\documentclass{jfm}
\usepackage{graphicx}
\usepackage{epstopdf, epsfig, bm}
\usepackage{color}

\shorttitle{Universality in statistics of Stokes flow over no-slip rough wall}
\shortauthor{V. Parfenyev, S. Belan and V. Lebedev}

\title{Universality in statistics of Stokes flow over no-slip wall with random roughness}
\author{Vladimir Parfenyev\aff{1,2}\corresp{\email{parfenius@gmail.com}}, Sergey Belan\aff{3}
    \and
    Vladimir Lebedev\aff{1,2}}

\affiliation{
\aff{1}Landau Institute for Theoretical Physics, 142432, Ak. Semenova 1-A, Chernogolovka, Russia
\aff{2}National Research University Higher School of Economics, 101000, Myasnitskaya 20, Moscow, Russia
\aff{3}Massachusetts Institute of Technology, Department of Physics, Cambridge, Massachusetts 02139, USA}

\begin{document}

\maketitle

\begin{abstract}
Stochastic roughness is widespread feature of natural surfaces and is an inherent by-product of most fabrication techniques. In view of rapid development of microfluidics, the important question is how this inevitable evil affects the low-Reynolds flows which are common for micro-devices. Moreover, one could potentially turn the flaw into a virtue and control the flow properties by means of specially "tuned" random roughness. In this paper we investigate theoretically the statistics of fluctuations in fluid velocity produced by the waviness irregularities at the surface of a no-slip wall.
Particular emphasis is laid on the issue of the universality of our findings.
\end{abstract}

\section{Introduction}\label{sec:introduction}

Effect of the wall texture on the fluid motion is a long-standing research topic of both fundamental and practical interest in a variety of fields.
Particularly important class of hydrodynamics problems concerns the role of random surface topography \citep{Darcy_1857,Nikuradse_1933,Moody_1944,Richardson_1973,Kandlikar_2003, Hao_2006, MacDonald_2017,Yan_2015, Kunert_2007,Vinogradova_2011, Zayernouri_2013}.
Given a sufficient spatial resolution, virtually any natural or manufactured surface exhibits irregular fluctuations of shape.
There is a need for understanding how the inevitable imperfectness affects the performance of fluidic devices and flow properties in natural systems.
This is particularly relevant to the low-Reynolds flows typical in nano/microfluidic and chemical micro-process applications where small length scales of devices imply significant boundary effects \citep{Taylor_2006,Chen_2009}.
One of the challenges here is that the statistical properties of the random surface roughness may be poorly specified or completely unknown a-priory.
Depending on the processing methods and material properties, resulted roughness of the manufactured surface may be Gaussian or non-Gaussian in height distribution, exponential or non-exponential in the decay of spatial correlations, isotropic or anisotropic, etc. Given such a diversity of alternatives, the issue of universality comes to the fore.
Can we formulate any general conclusions concerning the influence of random roughness on a fluid motion that hold true regardless of the fine details of the roughness statistics?

Here we implement a perturbative method together with statistical tools to address this question for the low-Reynolds flow of incompressible Newtonian fluid bounded by a no-slip rough wall. Clearly, any deviation of the boundary from ideal geometrical shape produces perturbation of the fluid velocity field anticipated in the idealized case. For the steady fluid motion in spatial domain with stochastically rough boundary, the resulting perturbation is stationary in time but fluctuates in space. Our analysis is focused on the statistical properties of these roughness-induced fluctuations in the "wavy" limit, when the characteristic height $h$ of the surface imperfections is much smaller than their characteristic transversal size $l \gg h$.

Let us briefly overview the structure and the main results of this paper.
Starting with the simplest setting, namely, the 2d and 3d flows in a half-space over a flat rough wall,
we characterize the fluid velocity fluctuations in terms of their spatial intensity profile, one-point probability distribution, pair correlation function and energy spectrum.
At the distances $z \gg l$  from the wall, these objects are found to exhibit universal behavior insensitive to the fine details of the roughness statistics. 
In particular, the far-field velocity fluctuations are Gaussian and their intensity as a function of distance from the wall demonstrates the power-law scaling.
Although the properties of the velocity fluctuations in the region $z\lesssim l$ are non-universal, it turns out that different surfaces produce flow patterns with common qualitative features.

Next, motivated by recent experiments and simulations by \citet{Jaeger_2012}, we consider the $2d$ and $3d$ fluid flows in a plane channel between two parallel walls, one of which is rough, while another is perfectly flat.
If the correlation length $l$ is much shorter than the distance between walls $d$, the statistical properties of the velocity fluctuations in the channel are universal.
Otherwise, when correlation length $l$ is comparable with $d$ or larger, the results are not universal, but again we pay a special attention to common qualitative features of the produced fluid flows.
The main difference between the fluid flow in the channel and in the half-space over a rough wall is that the large-scale velocity fluctuations are suppressed inside the channel. This is demonstrated by analysis of the energy spectrum of velocity fluctuations as a function of the distance to the rough wall in these two cases.

Finally, we study how the roughness-induced fluctuations in the stationary fluid flow affect the mass transport in the wall-normal direction. Being advected by the mean flow, the passive tracers are subject to the relatively weak flow disturbances generated by the rough wall. At sufficiently large time scale, one may expect to see the spatial dispersion of tracers due to these random disturbances. We show, however, that the spatial dispersion in the wall-normal direction quickly reaches some value and no longer changes with time/distance (along the channel). Therefore, the small-amplitude roughness of the channel walls does not lead to the mixing enhancement. 

\section{Model Formulation}

\begin{figure}
\centering\includegraphics[scale=.6]{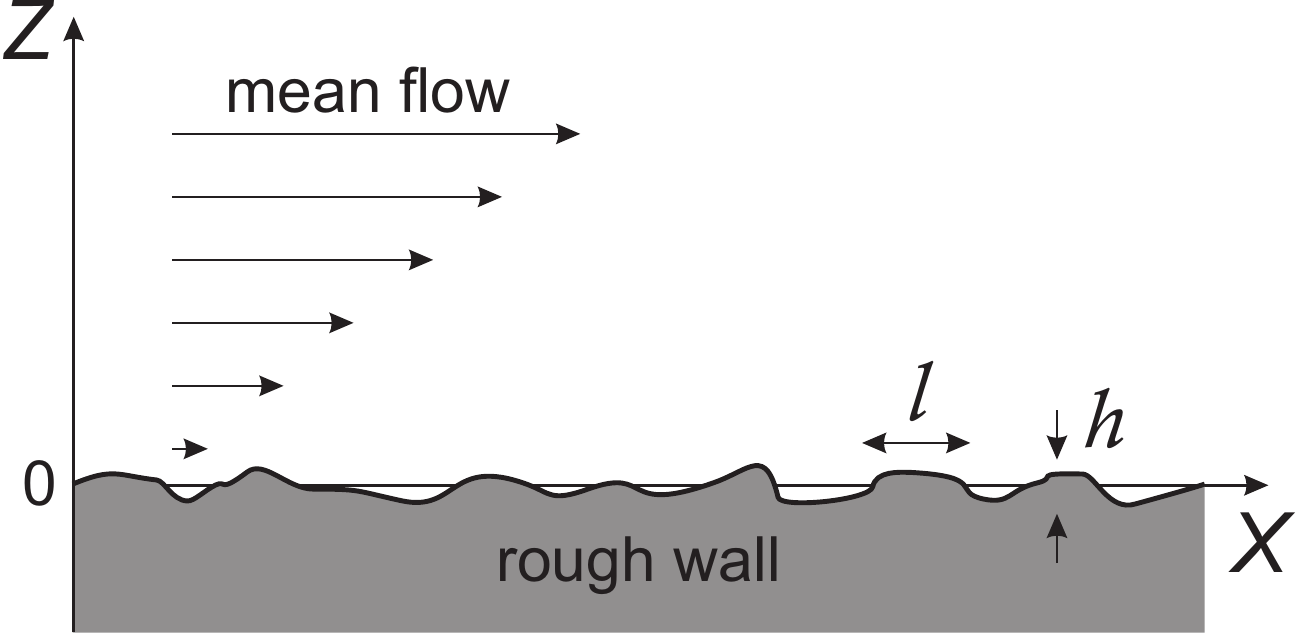}
\caption{Viscous fluid flow over a no-slip rough wall, which is flat in average but has random surface imperfections.}
\label{pic:setup}
\end{figure}

Let us consider the steady fluid flow in a half-space bounded by a wall which is flat in average but have random surface imperfections, see Fig.~\ref{pic:setup}. The $X$-axis of Cartesian coordinates is directed along the mean flow, while $Z$-axis is perpendicular to the wall. To illustrate the main ideas, first we use the two-dimensional model where wall texture is translationally invariant in the spanwise direction. The generalization to the three-dimensional case will be discussed in Sec.~\ref{sec:3d}. In the two-dimensional model the deviation of the wall surface from the average level $z = 0$ can be described as
\begin{equation}
z=\xi(x),
\end{equation}
where $\xi(x)$ is a single-valued function of the streamwise coordinate.

Within the statistical framework, the roughness amplitude $\xi(x)$ is characterized by the surface height distribution $P(\xi) = \langle \delta(\xi - \xi(x)) \rangle$, and the spatial coherence of the surface profile is described by the pair correlation function $F(r)=\langle\xi(x)\xi(x+r) \rangle$. Here the angle brackets mean the coarse-graining averaging over the scale, which is much larger than the correlation length $l$ of surface imperfections, but much smaller than the size of the system. We also imply homogeneity of the roughness statistics so that the height distribution $P(\xi)$ is the same at any point of the wall and the pair correlation function $F(r)$ is a function of inter-point distance $r$ and does not depend on the $x$-coordinate. Note that the function $F(r)$ is even in $r$.




While in most theoretical studies the surface height distribution is assumed to be Gaussian, experimental data indicate that real surfaces quite often exhibit non-Gaussian fluctuations in shape \citep{Ren_2011}. As for the pair correlation function of $\xi(x)$, it is believed to be well-approximated by exponential function for the most fabrication techniques \citep{Bhushan_2001,Ren_2011}. Being interested in the model independent features, here we do not specify the form of functions $P(\xi)$ and $F(r)$, assuming only that
the surface height distribution $P(\xi)$ has zero mean and finite variance, and the characteristic height of the surface imperfections $h$ is much smaller than their correlation length $l \gg h$. 
The first assumption ensures that the  surface asperities have well-defined characteristic amplitude $h$, while the second assumption means that we deal with slowly varying boundary, whose effect on the fluid motion
can be analyzed within the perturbative approach \citep{VanDyke_1964}, see Sec.~\ref{sec:perturb}.

When the Reynolds number is very small, the Stokes approximation is adequate in describing the fluid motion \citep{LL_fm}.
The Stokes flow of incompressible fluid is described by the equations
\begin{equation}
\eta\nabla^2 \vec v =\nabla p, \quad
\nabla \cdot \vec v=0,
\label{vis0}
\end{equation}
where $\vec v$ is the two-dimensional flow velocity, $p$ is pressure and $\eta$ is the dynamic viscosity coefficient. Note that $\nabla^2 p=0$, as a consequence of Eq. (\ref{vis0}). Next, we need to impose the no-slip boundary condition at the surface of the wall
 \begin{equation}
\label{bc0}
\vec v|_{z= \xi(x)}=0.
\end{equation}

\section{Perturbative Approach}
\label{sec:perturb}

In general, the fluid velocity field can be represented as
\begin{equation}\label{eq:velocity}
\vec v=\vec U+\vec u,
\end{equation}
where $\vec U$ is the velocity in the problem with perfectly flat wall and $\vec u$ is the roughness-induced correction.
Assuming that $\xi(x)=0$, we readily obtain from Eqs. (\ref{vis0}) and (\ref{bc0}) the shear velocity profile: $U_x=s z$, where $s$ is the shear rate. If the surface imperfections have characteristic height $h$ and transversal size $l\gg h$, then in analyzing the flow pattern in the region $z \gg h$ one can treat $\vec u$ as a relatively small perturbation to $\vec U$ obeying the following boundary value problem:
\begin{eqnarray} \label{vis1}
 &\eta\nabla^2 \vec u =\nabla p, \quad \nabla \cdot \vec u=0,\\
 \label{bc1}
 &u_x|_{z=0}=-s\xi(x), \ \ u_z|_{z=0}=0, \ \ \vec u|_{z \rightarrow \infty} \rightarrow 0.&
 \end{eqnarray}
Equation (\ref{bc1}) was obtained from the exact boundary condition (\ref{bc0}) by expanding the velocity field $\vec v$ near $z=0$ in a Taylor series. The condition $l \gg h$ of a small amplitude of roughness allows us to neglect all the other terms of this Taylor series.

The presented boundary perturbation method is  well-known in fluid mechanics \citep{VanDyke_1964} and has been widely used to analyze the effect of  small amplitude boundary roughness on the fluid motion, see, e.g.,  \cite{Lessen_1976,Wang_1976,Wang_1978,Wang_1979,PhanThien_1980,PhanThien_1981a,PhanThien_1981b,Pozrikidis_1987,Wang_2005,Wang_2006}.
Note, however, that in majority of these studies the roughness is considered to be regular: the surface profile is described by a single harmonic function or by a superposition of several harmonics. The only exception to this trend is a series of publications by \citet{PhanThien_1980,PhanThien_1981a,PhanThien_1981b}, where Stokes flow rate enhancement as a function  of  the statistics of the corrugation is analyzed. To the best our knowledge, our work is the first attempt to combine perturbative approach with the statistical tools to make analytical predictions concerning the statistics of flow fluctuations generated by stochastic roughness.

Due to the incompressibility condition (\ref{vis1}) the velocity field $\vec u=(u_x,u_z)$ can be written in terms of the stream function $\psi(x,z)$
 \begin{equation}
u_x=\partial_z\psi, \ \ \ u_z = -\partial_x\psi.
\end{equation}
By substituting these expressions into the boundary value problem (\ref{vis1})-(\ref{bc1}), we find that the stream function satisfies the biharmonic equation
\begin{equation}
\label{eq:biharmon}
\nabla^4 \psi (x,z) = \frac{\partial^4 \psi}{\partial x^4} + 2 \frac{\partial^4 \psi}{\partial x^2 \partial z^2}+ \frac{\partial^4 \psi}{\partial z^4}= 0,
\end{equation}
supplemented by the following boundary conditions
\begin{equation}
\label{eq:stream_bc}
\partial_z\psi|_{z=0}=-s\xi(x), \ \ \partial_x\psi|_{z=0}=0, \ \ \partial_z\psi|_{z=\infty}=\partial_x\psi|_{z=\infty}=0.
\end{equation}
Due to the linearity of Eq. (\ref{eq:biharmon}), the stream function $\psi(x,z)$ in the flow domain can be represented as an integral over its boundary distribution which is given by Eq. (\ref{eq:stream_bc}),
\begin{equation}
\label{stream_function}
\psi(x,z)=-\frac{sz^2}{\pi}\int\limits_{-\infty}^{+\infty}\frac{\xi(x')dx'}{(x-x')^2+z^2} + C,
\end{equation}
where $C$ is an arbitrary constant. Then, for the velocity field we find
\begin{equation}
\label{u_2d}
\displaystyle u_x  =-\frac{2sz}{\pi}\int\limits_{-\infty}^{+\infty}\frac{(x-x')^2\xi(x')dx'}{((x-x')^2+z^2)^2}, \quad u_z =-\frac{2sz^2}{\pi}\int\limits_{-\infty}^{+\infty}\frac{(x-x')\xi(x')dx'}{((x-x')^2+z^2)^2}.
\end{equation}

\section{Intensity of Velocity Fluctuations}

This result allows us to extract information regarding the statistical properties of the fluid flow. Let us first consider the single-point statistical moments of flow fluctuations. Since $\langle\xi\rangle=0$ by an assumption, then $\langle\vec u\rangle=0$. For the second moments we find
\begin{eqnarray}
\label{covariance_xz_2d}
&\displaystyle \langle u_x u_z \rangle = -\frac{4 s^2 z^3}{\pi^2} \int\limits_{-\infty}^{+\infty}  \int\limits_{-\infty}^{+\infty} \frac{dx' dx'' F(x''-x') x'^2 x''}{(x'^2 + z^2)^2 (x''^2 + z^2)^2},&\\
\label{covariance_xx_2d}
&\displaystyle \langle u_x^2\rangle=\frac{4 s^2 z^2}{\pi^2} \int\limits_{-\infty}^{+\infty}  \int\limits_{-\infty}^{+\infty} \frac{dx' dx'' F(x''-x') (x'x'')^2}{(x'^2 + z^2)^2 (x''^2 + z^2)^2},&\\
\label{covariance_zz_2d}
&\displaystyle \langle u_z^2\rangle = \frac{4 s^2 z^4}{\pi^2} \int\limits_{-\infty}^{+\infty}  \int\limits_{-\infty}^{+\infty} \frac{dx' dx'' F(x''-x') x'x''}{(x'^2 + z^2)^2 (x''^2 + z^2)^2}.&
\end{eqnarray}
The covariance $\langle u_x u_z \rangle$ of the fluctuating flow components is always zero because the correlation function $F(r)$ is even in $r$, but to calculate the second moments $\langle u_x^2\rangle$ and $\langle u_z^2\rangle$ one needs to know the particular form of the function $F(r)$. However, it is intuitively clear that the answer must become insensitive to statistics of the surface imperfectness when $z \gg l$.
Being far from the boundary we can assume that the random heights of surface asperities at different points of the wall are completely independent of each other and replace the autocorrelation function $F(x'-x'')$ by $A\delta(x'-x'')$, where
\begin{equation}
\label{eq:A}
A = \int\limits_{-\infty}^{+\infty}F(r)dr.
\end{equation}
This is a mere mathematical idealization which is, however, well justified when the roughness correlation length $l$ is small compared to the distance from the wall.
In result, we find that in the "far field" the fluctuation intensity is inversely proportional to the distance to the wall
\begin{equation}
\label{covariance}
\langle u_x^2\rangle=\langle u_z^2\rangle=\frac{As^2}{4\pi z}, \quad \langle u_x u_z\rangle=0.
\end{equation}
Note that the quantity $A$ can be estimated as $h^2 l$, where $h$ is the roughness height of the surface and $l$ is its correlation length.

It is worth discussing the difference between the effect of the stochastic roughness and that of the deterministic periodic waviness. For $\xi(x)=h\cos kx$ with $kh\ll1$ our perturbative scheme readily yields for $z\gg h$
\begin{equation}\label{eq:periodic}
  \langle u_x^2\rangle = \frac{s^2 h^2}{2} \left(|k|z-1 \right)^2 e^{-2|k|z}, \quad \langle u_z^2\rangle = \frac{s^2 h^2}{2} (kz)^2 e^{-2|k|z}, \quad \langle u_x u_z\rangle=0.
\end{equation}
We thus obtain an exponential decay of perturbation produced by deterministic waviness. In contrast, the random roughness exhibits the long-ranged (scale-free) effect.

\section{Statistics of Velocity Fluctuations}
\label{sec:pdf}
More generally, one could be interested in the full probability density function ${\cal P}(\vec u)$ of flow fluctuations at a given distance $z$ from the wall. The characteristic function of this probability distribution is defined as its Fourier transform
\begin{eqnarray}
{\cal G}(\vec k)=\int\limits_{-\infty}^{+\infty} e^{i\vec k \vec u}{\cal P}(\vec u) du_x du_z.
\end{eqnarray}
Replacing averaging over space by averaging over ensemble and using the technique of functional integration, one obtains the following result for $z \gg l$ (see Appendix~\ref{appA})
\begin{equation}\label{char_function}
{\cal G}(\vec k)= \exp \left( -\frac{A s^2 k^2}{8 \pi z} \right),
\end{equation}
which after the inverse Fourier transform gives the Gaussian distribution with zero mean vector
\begin{equation}
\label{PDF1}
P(\vec u)=\frac{1}{\sqrt{(2\pi)^2 \det \hat\Sigma}}\exp \left( -\frac{1}{2}\vec u^T\hat\Sigma^{-1}\vec u \right).
\end{equation}
The elements of covariance matrix $\Sigma_{ij}=\langle u_iu_j\rangle$ are defined by Eq. (\ref{covariance}).

Thus, the far-field fluctuations are Gaussian as long as the surface height distribution $P(\xi)$ has finite variance $\langle \xi^2\rangle=F(0)$.
Essentially this result is a direct manifestation of universality proclaimed by the famous central limit theorem for the sum of large number of independent random variables.
Indeed, as we see from Eq. (\ref{u_2d}),  the velocity field is given by the weighted superposition of the contributions produced by different points of the rough boundary.
At $z\gg l$, the answer is determined mainly by the sum of $\sim z/l$ independent random terms corresponding to the nearest surface asperities, and, thus, the central limit theorem comes into play.



\section{Pair Correlation Tensor and Energy Spectrum}

So far we have discussed the single-point statistics of the fluid velocity field. To get insight into the flow structure, let us consider a correlation tensor of velocity fluctuations
\begin{equation}
R_{ij}(r,z_1,z_2)=\langle u_i(x,z_1)u_j(x+r,z_2)\rangle.
\end{equation}
Using Eqs. (\ref{u_2d}) we find the components of this tensor in Fourier space
\begin{eqnarray}
\label{Rzz2d_0}
&\tilde R_{zz}(k,z_1,z_2)=s^2 z_1z_2 k^2 \tilde F(k)e^{-(z_1+z_2)|k|},&\\
\label{Rxx2d_0}
&\tilde R_{xx}(k,z_1,z_2)=s^2(z_1|k|-1)(z_2|k|-1)\tilde F(k)e^{-(z_1+z_2)|k|},&\\
\label{Rxz2d_0}
&\tilde R_{xz}(k,z_1,z_2)= i s^2 (z_1 |k|-1)z_2 k \tilde F(k)e^{-(z_1+z_2)|k|}.&
\end{eqnarray}
Here $\tilde R_{ij}(k,z_1,z_2)=\int_{-\infty}^{+\infty}R_{ij}(r,z_1,z_2)e^{-ikr}dr$ and $\tilde F(k)=\int_{-\infty}^{+\infty}F(r)e^{-ikr}dr$.

At $z_1,z_2\gg l$ one can insert $\tilde F(k) \approx \tilde F(0)=A$.
This leads to the following universal result for the far-field correlation tensor:
\begin{eqnarray}
\label{Rzz2d}
&R_{zz}(r,z_1,z_2)=\displaystyle\frac{2As^2}{\pi}\frac{z_1z_2(z_1+z_2)[(z_1+z_2)^2-3r^2]}{[(z_1+z_2)^2+r^2]^3},&\\
\label{Rxx2d}
&R_{xx}(r,z_1,z_2)=\displaystyle\frac{2As^2}{\pi}\frac{z_1z_2(z_1+z_2)^3+(z_1^3+z_2^3)r^2+(z_1+z_2)r^4}{[(z_1+z_2)^2+r^2]^3},&\\
\label{Rxz2d}
&R_{xz}(r,z_1,z_2)=\displaystyle\frac{2As^2}{\pi}\frac{z_2 r [(2z_1+z_2)r^2 - (2z_1-z_2)(z_1+z_2)^2]}{[(z_1+z_2)^2+r^2]^3}.&
\end{eqnarray}
Thus, the roughness-induced fluctuations of fluid velocity exhibit algebraic decay of correlation in both streamwise and wall-normal directions.

A particularly important characteristic of the flow at distance $z$ from the wall is the energy spectrum of velocity fluctuations, which is defined as
\begin{equation}
\label{eq:spectr2d}
E(k)=\frac{1}{4\pi}(\tilde R_{xx}(k,z,z)+\tilde R_{zz}(k,z,z)).
\end{equation}
From Eqs. (\ref{Rzz2d_0}-\ref{Rxx2d_0}) we readily obtain
\begin{equation}
\label{eq:spectrum}
E(k)= \frac{s^2}{4\pi}\tilde F(k)[(z|k|-1)^2+z^2k^2]e^{-2z|k|}.
\end{equation}
The energy spectrum and the correlation length of roughness-induced fluctuations have been measured for the low-Reynolds flow in channel between two parallel plates in \cite{Jaeger_2012}. However, in this work the channel width is of the order of the roughness correlation length. Thus, the lack of sufficient experimental data does not allow us to test the already obtained predictions. We can only declare that the experimental data confirm that at low flow rates (low Reynolds number) the flow structure is determined by roughness characteristic of the surface and is insensitive to flow rate. In Sec.~\ref{sec:channel} we extend our model to the case of a plane channel and then we carry out a more detailed comparison between theory and experiment.

\section{Near-Wall Velocity}
\label{sec:NF}

\begin{figure}
\centering\includegraphics[width=\linewidth]{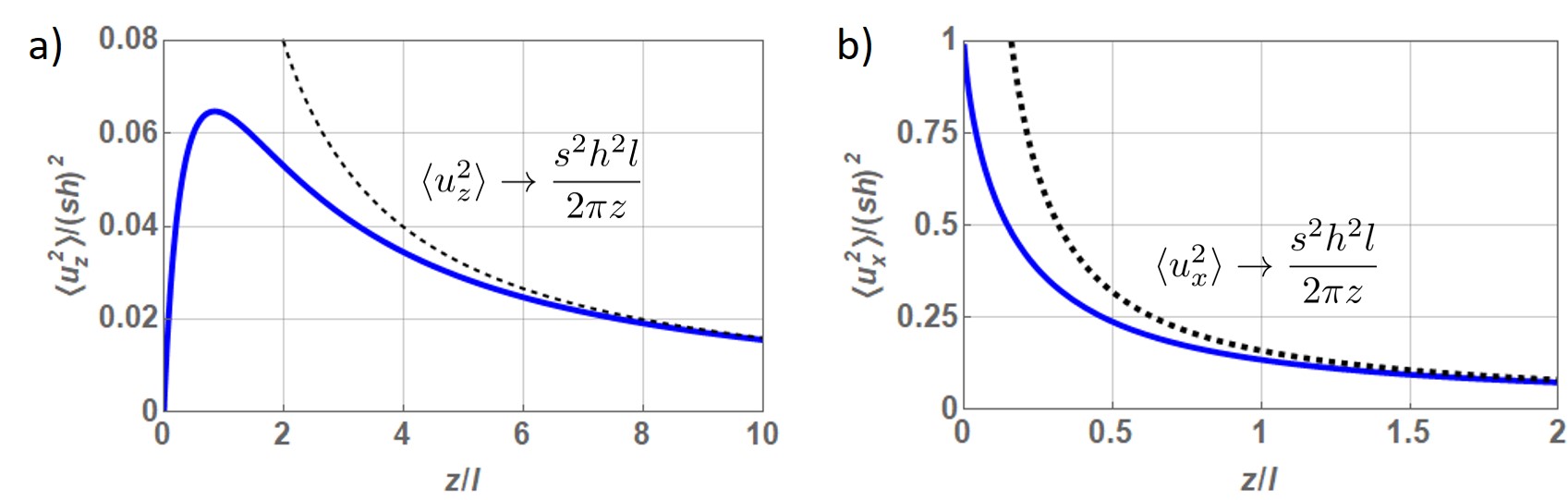}
\caption{The wall-normal profile of fluctuation intensities $\langle u_z^2\rangle$ and $\langle u_x^2\rangle$} for the exponential surface autocorrelation function $F(r)=h^2 e^{-|r|/l}$. The dashed lines represent the asymptotic dependence (\ref{covariance}) at $z \gg l$.
\label{pic:intensity}
\end{figure}

As mentioned above, the near-field velocity is sensitive to the roughness structure. However, as it is clear from Eqs. (\ref{covariance_xz_2d})-(\ref{covariance_zz_2d}), various surfaces produce flow pattens with common qualitative features.
In particular, the intensity $\langle u_z^2 \rangle$ of fluctuations in wall-normal velocity as a function of distance from the wall is always non-monotonic, taking its maximum at distance of the order of the roughness correlation length $l$, see Fig.~\ref{pic:intensity}a for illustration. In contrast, the dependence of fluctuation intensity $\langle u_x^2 \rangle$ on the distance from the wall is usually monotonic and then it takes its maximum value on the wall, see Fig.~\ref{pic:intensity}b. Interestingly, the locations of these maximum values are completely independent of the characteristic height $h$ of surface imperfection. On the other hand, the maximum values do not depend on $l$, but are only determined by $h$.

This conclusion agrees well with those reported before. For example, in the paper \citep{Zayernouri_2013} the Stokes flow in a channel with rough wall was solved numerically and it was found that location of maximum of fluctuations in wall-normal velocity moves toward the rough wall as correlation length decreases and is insensitive to the roughness’ variance.
Similar features have been  observed for the deterministic roughness in the form of sinusoidal waviness for the channel flows with moderate Reynolds number, see \citet{Wang_2005,Wang_2007}. Namely, the increase of the wall wavenumber is shown to result in decrease of the disturbance area of the flow field, while the relative roughness determines the maximum disturbance amplitude, but does not affect the width of disturbed area. Note also that in the vicinity of the wall the streamwise component of velocity predominates in the intensity of velocity fluctuations, $\langle u_x^2 \rangle \gg \langle u_z^2 \rangle$, while in the far-field region, $z \gg l$, both components are equal to each other, see expression (\ref{covariance}).

As for the spatial decay of correlations in flow fluctuations, the velocity correlation length in the region $z\lesssim l$ is of the order of $l$. The energy spectrum of fluctuations (\ref{eq:spectrum}) at $h \ll z \lesssim l$ is determined by the power spectral density function of surface roughness $\tilde F(k)$, which is the Fourier transform of the surface autocorrelation function $F(r)$. Say, for exponential correlator $F(r)=h^2 e^{-|r|/l}$, that is typical for most etched surfaces, we have $\tilde{F}(k)=2h^2 l /(k^2 l^2+1)$ and, therefore, the near-field spectrum displays the power-law with exponent equal to $-2$ in the region $1/z \gg k \gg 1/l$ and decays exponentially when $k \gg 1/z$.

The dependence of velocity fluctuations on the roughness statistics in the near-wall region allows us to propose the hydrodynamic method of investigation of surface properties:  measuring the energy spectrum of velocity fluctuations $E(k)$ and using expression (\ref{eq:spectrum}), one can infer the power spectral density function $\tilde{F}(k)$ of the surface.

\section{3D Flow Over Flat Surface}
\label{sec:3d}

The perturbative approach can be applied to the fluctuations in three-dimensional flow produced by an imperfect wall whose surface $z=\xi(x,y)$ is stochastically rough in both streamwise and spanwise directions. Now the spatial velocity fluctuations are given by \citep{jansons1988general}
\begin{eqnarray}
\label{ux3d}
u_x(x,y,z)=-\frac{3sz}{2\pi}\int\limits_{-\infty}^{+\infty}\int\limits_{-\infty}^{+\infty}
\frac{dx' dy'(x-x')^2\xi(x',y')}{((x-x')^2+(y-y')^2+z^2)^{5/2}},\\
\label{uy3d}
u_y(x,y,z)=-\frac{3sz}{2\pi}\int\limits_{-\infty}^{+\infty}\int\limits_{-\infty}^{+\infty}
 \frac{dx' dy'  (x-x')(y-y') \xi(x',y')}{((x-x')^2+(y-y')^2+z^2)^{5/2}}, \\
 \label{uy3d}
u_z(x,y,z)=-\frac{3sz^2}{2\pi}\int\limits_{-\infty}^{+\infty}\int\limits_{-\infty}^{+\infty} \frac{dx' dy'(x-x') \xi (x',y')}{((x-x')^2+(y-y')^2+z^2)^{5/2}}.
\end{eqnarray}
Note that if the surface elevation $\xi(x,y)$ does not depend on $y$-coordinate then the integration over the spanwise direction can be performed and one obtains the previously found result (\ref{u_2d}). Further we will assume that the roughness statistics is homogeneous and isotropic.

In the far-field region, $z \gg l$, the velocity fluctuations have the Gaussian probability density function $P(\vec u)=((2\pi)^3 \det \hat\Sigma)^{-1/2}\exp \left( -\frac{1}{2}\vec u^T\hat\Sigma^{-1}\vec u \right)$
with covariance matrix $\Sigma_{ij}=\langle u_i u_j\rangle$ given by
\begin{eqnarray}\label{eq:covariance_3d}
&\displaystyle \langle u_x^2\rangle=\frac{9As^2}{128\pi z^2}, \quad \langle u_y^2\rangle=\frac{3As^2}{128\pi z^2}, \quad \langle u_z^2\rangle=\frac{3As^2}{32\pi z^2},&\\
&\displaystyle \langle u_xu_y\rangle= \langle u_xu_z\rangle= \langle u_yu_z\rangle=0.&
\end{eqnarray}
Now the quantity $A$ is estimated as $h^2l^2$, where as before $h$ and $l$ denote the roughness height of the surface and its correlation length, respectively.

The velocity correlation tensor is defined by $R_{ij}(\vec r,z_1,z_2) = \langle u_i(\vec 0,z_1)u_j(\vec r, z_2)\rangle$ where $\vec r=(x,y)$ is the vector in $x-y$ plane.
As in the two-dimensional case, velocity correlations at $z_1,z_2 \gg l$ exhibit the algebraic decay in streamwise, spanwise, and wall-normal directions. 
The energy spectrum $E(\vec k)$ of velocity fluctuations is determined by the trace of the correlation tensor in Fourier space, namely
\begin{equation}
\label{eq:spectr3d}
E(\vec k)=\frac{1}{8\pi^2} \left( \tilde R_{xx}(\vec k,z,z)+\tilde R_{yy}(\vec k,z,z)+\tilde R_{zz}(\vec k,z,z) \right),
\end{equation}
where $\tilde R_{ij}(\vec k,z_1,z_2) = \int_{-\infty}^{+\infty}R_{ij}(\vec r,z_1,z_2)e^{-i\vec k \vec r}d^2 \vec r$ and $\vec k = (k_x, k_y)$.
Using Eqs. (\ref{ux3d}-\ref{uy3d}) one finds $\tilde{R}_{xx} (\vec{k},z,z) = s^2 \tilde{F}(k) \left( 1 - \frac{k_x^2}{k} z \right)^2 e^{-2kz}$, $\tilde{R}_{yy} (\vec{k},z,z) = s^2 \tilde{F}(k) \left( \frac{k_x k_y z}{k} \right)^2 e^{-2kz}$, $\tilde{R}_{zz} (\vec{k},z,z) = s^2 \tilde{F}(k) k_x^2 z^2 e^{-2kz}$, where we introduced $k = |\vec k|$, and then
\begin{equation}
\label{eq:spectr3d_calc}
E(\vec k)= \frac{s^2}{8 \pi^2} \tilde F(k) \left( 1 - 2 \frac{k_x^2}{k}z + 2 k_x^2 z^2 \right) e^{-2kz}.
\end{equation}
Note that by substituting $k_y=0$, we can restore the previously obtained answer (\ref{eq:spectrum}) for the two-dimensional problem.

In experimental papers, the energy spectrum $E_{av} (k)$ averaged over the direction of the vector $\vec k$ is often calculated, 
\begin{equation}
\label{eq:spectr_av}
E_{av} (k) = k \int\limits_{0}^{2 \pi} d \varphi \, E (\vec k),
\end{equation}
where the integration is performed over the polar angle in the $k_x-k_y$ plane. By substituting expression (\ref{eq:spectr3d_calc}) into equation (\ref{eq:spectr_av}), we obtain
\begin{equation}
\label{eq:spectr_3d_av}
E_{av} (k) = \frac{s^2}{4 \pi} \tilde F(k) (1-kz+k^2 z^2) k e^{-2kz}.
\end{equation}
In the case of the exponential surface autocorrelation function $F(r)=h^2 e^{-|r|/l}$, one has $\tilde F(k)=2\pi h^2 l^2/(k^2l^2+1)^{3/2}$ and, therefore, the near-field averaged over angle spectrum $E_{av} (k)$ displays the power-law tail with exponent equal to $-2$ in the region $1/z \gg k \gg 1/l$ and decays exponentially when $k \gg 1/z$.

\section{Flow Between Two Plates}
\label{sec:channel}

\begin{figure}
\centering\includegraphics[scale=.6]{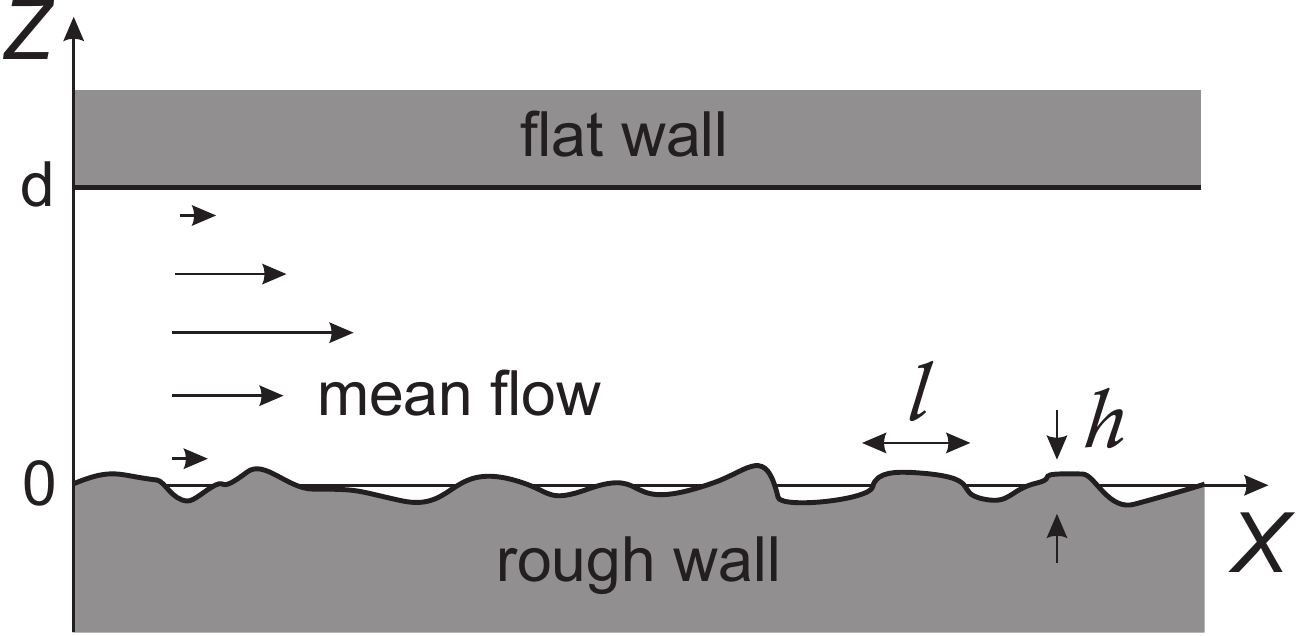}
\caption{Viscous fluid flow between no-slip rough wall and no-slip perfectly flat wall.}
\label{pic:setup2}
\end{figure}

In many practical situations the correlation length of surface irregularities is of the order of channel diameter so that the roughness cannot be treated as shortly-correlated.
What can one say about interplay between the roughness statistics and flow behaviour in this case? The scheme presented here allows to address the questions of this kind as well.
Motivated by the experiment \citep{Jaeger_2012}, in this section we consider the Stokes flow in a plane channel of width $d$. The bottom wall of the channel, which corresponds to $z=\xi(x)$, is rough, while the top wall, $z=d$, is perfectly flat, see Fig.~\ref{pic:setup2}. First, we consider two-dimensional model, the generalization to three dimensions is straightforward.

\subsection{2D plane channel}

The fluid flow in the channel is created by applying the pressure gradient $\nabla P_0$ along $X-$direction. The solution of the problem without roughness is well-known Poiseuille flow
\begin{equation}
\label{eq:Poiseuille}
U_x(z) = s z (1-z/d), \quad s = - \frac{\nabla P_0}{2\eta} d,
\end{equation}
and then the roughness-induced velocity correction $\vec u$ must satisfy the same boundary value problem (\ref{vis1})-(\ref{bc1}), but with the additional no-slip condition $\vec u\vert_{z=d} = 0$, which is posed on the top wall. Note again that the used perturbative approach is correct only in the limit of small-amplitude roughness, when the characteristic height $h$ of the surface imperfections is much smaller than their characteristic transversal size $l \gg h$.

The considered boundary value problem can be solved in the Fourier space, $\tilde{f} (q_x, z) = \int_{-\infty}^{+\infty} dx \, f (x,z) e^{-i q_x x}$, and then for the velocity fluctuations one obtains
\begin{eqnarray}
\nonumber
\displaystyle \tilde{u}_x = \frac{s \tilde{\xi}}{1+ 2 d^2 q^2- \cosh (2 d q)} && \Big[ \cosh (q (2 d-z)) + \left[ 2 d q^2 (z-d)-1\right] \cosh (q z)\\
&&- q z \sinh (q (2 d-z))+ q (2 d-z) \sinh (q z) \Big],\\
\nonumber
\displaystyle \tilde{u}_z =\frac{-i q_x s \tilde{\xi}}{1 + 2 d^2 q^2-\cosh (2 d q)} && \Big[ z \cosh (q (2 d-z)) -z \cosh (q
   z) \\
&&+ 2 d q (z-d) \sinh (q z) \Big],
\end{eqnarray}
where we denote the wave number by $q = |q_x|$. The important characteristic of the fluid flow is the energy spectrum of velocity fluctuations $E(k)$, which was defined by expression (\ref{eq:spectr2d}). Now it is equal to
\begin{equation}
E(k) = \frac{1}{8 \pi^2} \int\limits_{-\infty}^{+\infty} dq \,  \left\langle \tilde{u}_x(q) \tilde{u}_x(k) + \tilde{u}_z(q) \tilde{u}_z(k) \right\rangle,
\end{equation}
and by using $\langle \tilde{\xi}_q \tilde{\xi}_k \rangle = 2\pi \delta(k+q) \tilde{F}(k)$, we find
\begin{eqnarray}
\nonumber
E(k) &=& \frac{s^2 \tilde F(k)}{4\pi \left[1 + 2 d^2 k^2-\cosh
   (2 d k)\right]^2}  \Bigg[ \Big( \cosh (k (2 d-z)) -k z \sinh (k (2d-z)) \\
\nonumber
&& + \left(2 d k^2 (z-d)-1\right) \cosh (k z) +k (2 d-z) \sinh (k z)\Big)^2 + k^2 \Big(-z \cosh (k z) \\
\label{eq:spectr2d_calc}
&& + 2 d k (z-d) \sinh (k z)+z \cosh (k (2 d-z)) \Big)^2 \Bigg].
\end{eqnarray}

\begin{figure}
\centering\includegraphics[width=\linewidth]{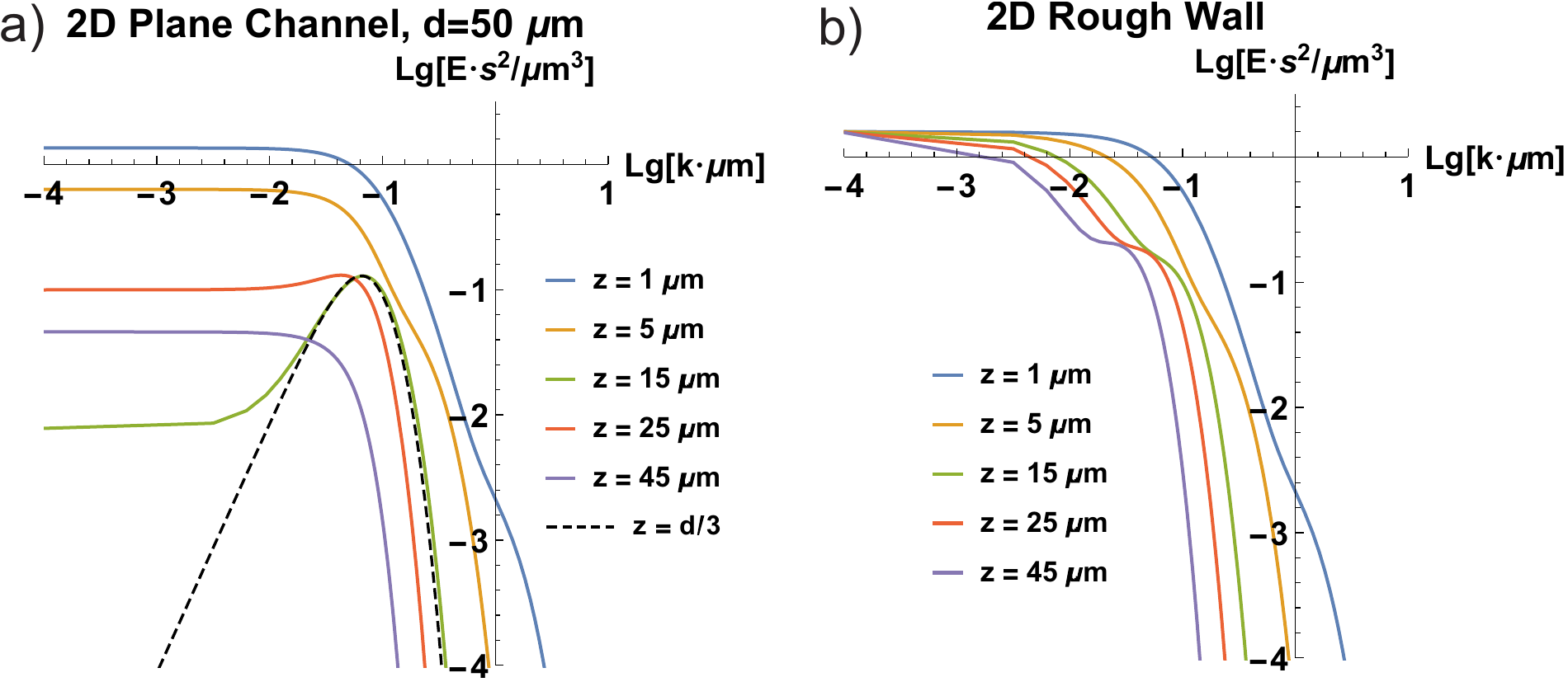}
\caption{(a) The energy spectrum (\ref{eq:spectr2d_calc}) in a channel of width $d=50\, \mu m$ and (b) The energy spectrum (\ref{eq:spectrum}) for the flow over the plane depending on the distance $z$ to the rough wall. In both cases the surface imperfections are correlated exponentially, $F(r)=h^2 e^{-|r|/l}$, with the parameter $l = 10 \, \mu m$. Numerical values correspond to the parameter $sh=1 \, \mu m/s$.}
\label{pic:spectr2d}
\end{figure}

To illustrate expression (\ref{eq:spectr2d_calc}), we consider the fluid flow in the plane channel of width $d=50 \, \mu m$, assuming that surface imperfections on the bottom wall have exponential correlation function $F(r)=h^2 e^{-|r|/l}$ with the parameter $l = 10 \, \mu m$. The dependence of the energy spectrum $E(k)$ of velocity fluctuations on distance $z$ to the bottom wall is presented in Fig.~\ref{pic:spectr2d}a. For comparison, Fig.~\ref{pic:spectr2d}b shows similar spectrum (\ref{eq:spectrum}) for the fluid flow over the rough wall.

Near the rough wall, $z \ll d$, both energy spectra look identical. The rough wall is able to generate fluctuation with $k \lesssim 1/l$ and this point corresponds to the kink in the spectra. The small bend on curves $z=1 \, \mu m$ in Figs.~\ref{pic:spectr2d}a and \ref{pic:spectr2d}b near point $\lg k =0$ corresponds to the change in asymptotic behaviour, which was discussed in Sec.~\ref{sec:NF}. In the region $k \gg 1/z$ one obtains $E(k) \propto e^{-2kz}$, while in the region $1/l \ll k \ll 1/z$ --- $E(k) \propto k^{-2}$. Far from the wall, when $z > l$, we observe only exponential decay.

As the distance to the rough wall increases, the energy spectra become different. First, the channel suppresses large-scale fluctuations, see curves $z=5 \, \mu m$. At distance $z \sim 0.2-0.5 \, d$ the spectrum in the channel becomes non-monotonic and exhibits maximum, see curve $z=15 \, \mu m$. The appearance of maximum and its position are weakly dependent on the correlation length $l \lesssim d$. Thus, the width $d$ of the channel imposes some restrictions on the size of fluctuations, which are able to penetrate inside the depth.

\begin{figure}
\centering\includegraphics[scale=.6]{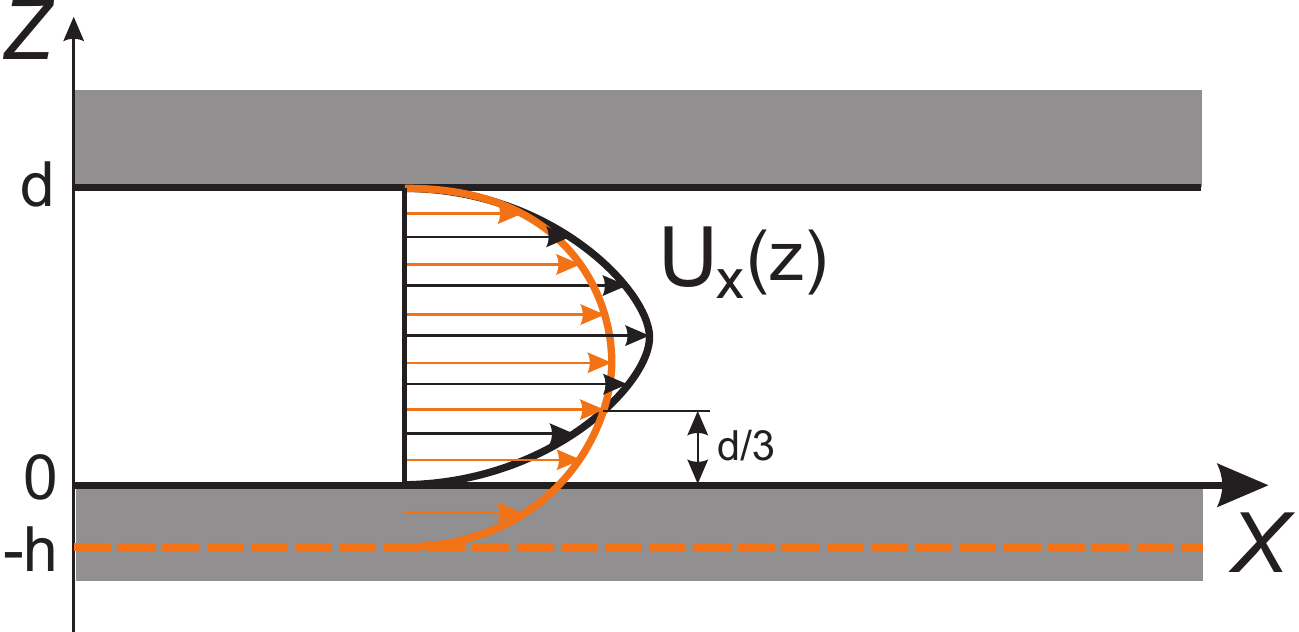}
\caption{Schematic picture, which explains the total suppression of large-scale velocity fluctuations in the channel at a distance $z=d/3$ from the rough wall. The black curve represents the velocity field (\ref{eq:Poiseuille}) for the problem without roughness, and the orange curve shows the velocity field (\ref{eq:Poiseuille2}) modified by large-scale surface imperfection.}
\label{pic:explanation}
\end{figure}

Interestingly, the large-scale velocity fluctuations in the channel are completely suppressed, i.e. $E(k) \rightarrow 0, \, k \rightarrow 0$, at a certain distance $z=d/3$ from the bottom wall, see dashed line in Fig.~\ref{pic:spectr2d}a. This result can be explained in the following way.
Due to the linearity of the problem, the large-scale velocity fluctuations are generated by the large-scale surface imperfections.
Clearly, a surface asperity, whose amplitude is $h$ and the streamwise size is much larger than $d$,  locally changes the channel width as shown in Fig.~\ref{pic:explanation}.
The used perturbative approach preserves the discharge of fluid in the linear approximation with respect to the surface elevation $h \ll d$.
Therefore, the velocity field for the channel with modified width is
\begin{equation}
\label{eq:Poiseuille2}
v_x(z) = s \frac{d^3}{(d+h)^3} (z+h) (1-z/d), \quad v_z = 0,
\end{equation}
where $z=-h$ corresponds to the new position of the bottom wall. The obtained velocity profile intersects with the solution of the problem without roughness (\ref{eq:Poiseuille}) in the point $z=d/3$, and therefore there is no large-scale velocity fluctuations at this distance from the rough wall.

Further increasing distance to the wall, one can see that the value $E(k)$ in the maximum is practically does not change, while the energy density of large-scale fluctuations increases by an order of magnitude, see curve $z=25 \, \mu m$ in Fig.~\ref{pic:spectr2d}a. After the maximum disappears, the spectrum gradually decreases for all wave numbers with increasing distance to the bottom wall, turning to zero on the top perfectly flat wall.

\subsection{3D plane channel}

Experimentally, the energy spectrum was measured in a channel between two parallel plates in \cite{Jaeger_2012}. However in that work the bottom plate was stochastically rough in both streamwise and spanwise directions. Our approach can be straightforwardly generalized to this case as well. To implement the direct comparison, now we consider the three-dimensional plane channel of width $d=51 \, \mu m$, and the correlation function of surface imperfections on the rough wall is assumed to be exponential, $F(r)=h^2 e^{-|r|/l}$, with the parameter $l = 12.9 \, \mu m$. Rms of surface fluctuations was equal to $h=19 \, nm \ll l$ and thus the perturbative approach is well justified.

\begin{figure}
\centering\includegraphics[width=\linewidth]{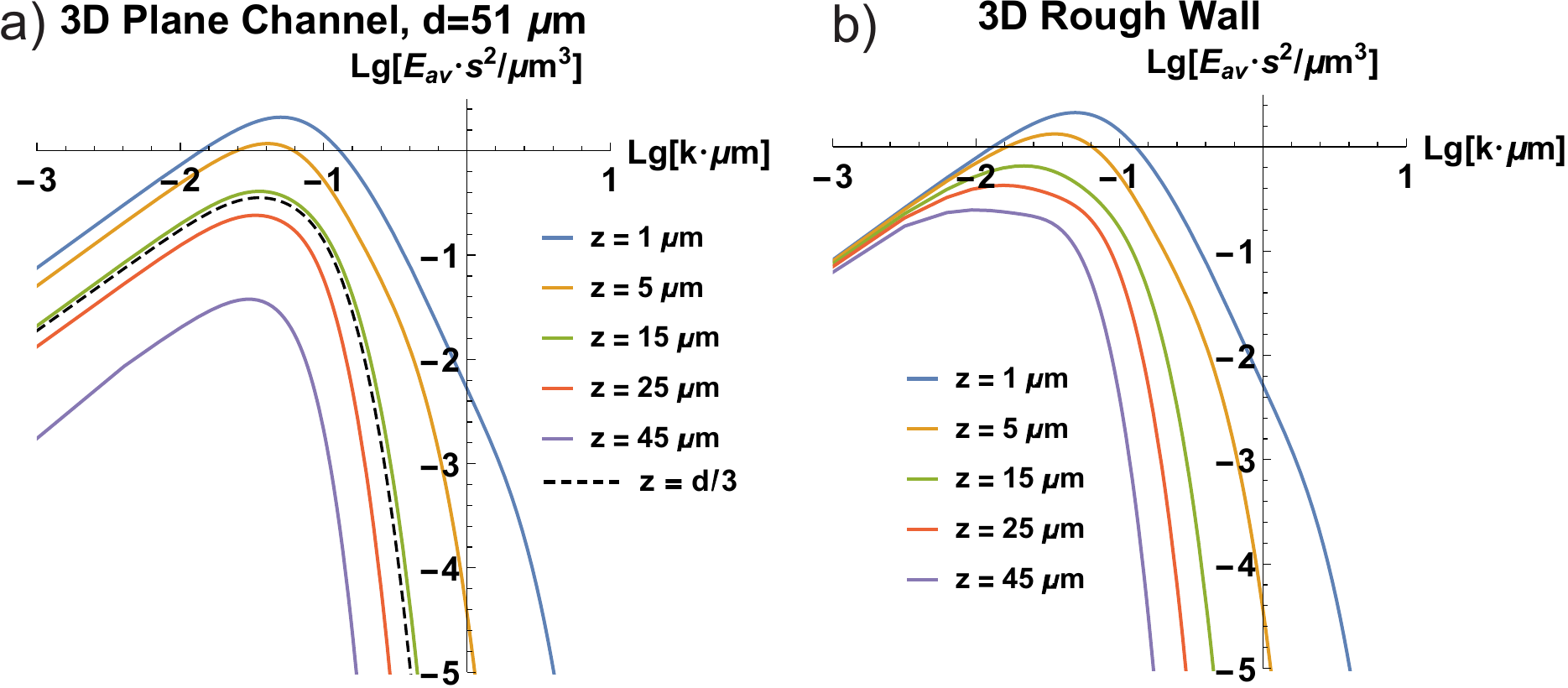}
\caption{(a) The energy spectrum $E_{av}(k)$ in a 3D channel of width $d=51\, \mu m$ and (b) The energy spectrum (\ref{eq:spectr_3d_av}) for the 3D flow over the plane depending on the distance $z$ to the rough wall. In both cases the surface imperfections are correlated exponentially, $F(r)=h^2 e^{-|r|/l}$, with the parameter $l = 12.9 \, \mu m$. Numerical values correspond to the parameter $sh=1 \, \mu m/s$.}
\label{pic:spectr3d}
\end{figure}

The energy spectrum of velocity fluctuations (\ref{eq:spectr_av}) in the 3D plane channel is shown in Fig.~\ref{pic:spectr3d}a for different distances $z$ to the rough wall. For comparison, Fig.~\ref{pic:spectr3d}b shows similar spectrum (\ref{eq:spectr_3d_av}) for the three-dimensional fluid flow over the rough plane. The large-scale behaviour, $k \rightarrow 0$, is described by $E_{av}(k) \propto k$ due to the phase volume factor, arising from averaging over the angle in the $k_x - k_y$ plane, see expression (\ref{eq:spectr_av}). One can see that the channel suppresses large-scale fluctuations as in the two-dimensional case, but now the distance $z=d/3$ does not play any special role, because the fluid flow has additional degree of freedom and the flow rate can depend on the $y$-coordinate. The small-scale behaviour demonstrates two regimes, as we have discussed in Sec.~\ref{sec:3d}. In the region $k \gg 1/z$ one observes $E_{av}(k) \propto e^{-2kz}$, while in the region $1/l \ll k \ll 1/z$ --- $E_{av}(k) \propto k^{-2}$. The last intermediate asymptotic can be observed only near the rough wall, see curves $z=1 \, \mu m$ in Fig.~\ref{pic:spectr3d}. Far from the wall, when $z > l$, the decay is exponential.

We must conclude that our theoretical findings are different from the results of experimental measurements, which are presented in \cite{Jaeger_2012}. The experimental spectrum is monotonic, while our theoretical analysis shows that the spectrum has a maximum, see Fig.~\ref{pic:spectr3d}a. Qualitatively, the large-scale behaviour demonstrates a reasonable agreement, while the small-scale behaviour is completely different. We would like to stress that predicted exponential decay $\propto e^{-2kz}$ of small-scale fluctuations is natural. Due to linearity of the problem, the velocity fluctuations with wave number $k$ are generated by the surface imperfections with the same wave number, and therefore their decay is exponential, see expression (\ref{eq:periodic}). We do not know how to explain the absence of an exponential decay in the experimental data. In part, the discrepancy between the theory and the measurements can be attributed to the speculation that the pair correlation function of surface $F(r)$ is not exponential, but has a more complex structure and depends on several scales. This speculation is based on the results of the experiment, in which the presence of several characteristic scales can be observed.





\section{Dispersion of Passive Scalar}

In this section we consider transport of the passive scalar in the steady velocity field with a relatively weak random component generated by the wall imperfections in the direction perpendicular to the wall. As above we write the fluid velocity as $\vec v=\vec U+\vec u$, where $\vec U = (U_x(z),0)$ represents the regular flow and $\vec u$ is the spatially random perturbation. We start with the two-dimensional fluid flow over the rough plane and then we will briefly discuss other considered geometries as well.
The passive tracer evolution is governed by the interplay of spatially varying mean flow and flow fluctuations
\begin{equation}\label{eq:scalar}
\partial_t \theta + (\vec{U} + \vec{u}) \nabla \theta = 0.
\end{equation}
Here we omitted the molecular diffusion term because we would like to focus on the mass transport produced only by the roughness-induced velocity fluctuations. A formal solution of Eq.~(\ref{eq:scalar}) is well known. The advection occurs along characteristics $\vec{\rho} (\vec{r},t)$ that have to be determined by solving the system of equations
\begin{equation}\label{eq:characteristic}
  \partial_t \vec{\rho} = \vec{U} (\vec{\rho}) + \vec{u}(\vec{\rho})
\end{equation}
with corresponding initial condition.

\begin{figure}
\centering\includegraphics[width=0.8\linewidth]{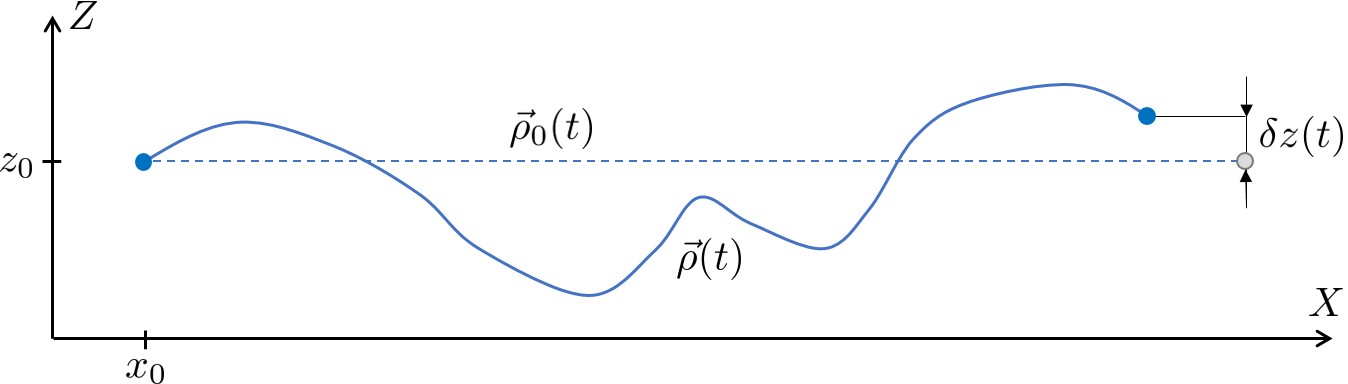}
\caption{The advection of the passive tracer along the characteristic $\vec{\rho}(t)$. The dashed line corresponds to the advection only by the mean flow $U_x(z)$.}
\label{pic:scalar}
\end{figure}

Let us find the displacement $\delta z(t)$ of the passive tracer in the wall-normal direction, assuming that initially it was located at some point $(x_0,z_0)$, see Fig.~\ref{pic:scalar}. Since the roughness-induced fluctuations are significantly smaller than the mean flow, we can solve Eq.~(\ref{eq:characteristic}) within the framework of the perturbation theory, i.e. $\vec{\rho} = \vec{\rho}_0 + \vec{\rho}_1 + \cdots$, where
\begin{eqnarray}
\partial_t \vec{\rho}_0 = \vec{U} (\vec{\rho}_0),&& \quad \vec{\rho}_0(0) = (x_0,z_0),\\
\partial_t \rho_{1i} = \partial_k U_i (\vec{\rho_0}) \rho_{1k} + u_i (\vec{\rho}_0),&& \quad \vec{\rho}_1(0) = (0,0).
\end{eqnarray}
The solution is $\vec{\rho}_0 = \left( x_0 + U_0 t, z_0 \right)$ where $U_0 = U_x(z_0)$, which corresponds to the advection by the mean flow, and for the first-order correction we find
\begin{equation}\label{eq:delta_z}
  \delta z (t) \approx \rho_{1z} (t) = \int\limits_0^t d \tau \, u_z (x_0 + U_0 \tau, z_0).
\end{equation}
Note that to calculate the second-order correction to $\delta z(t)$, one should not only find $\rho_{2z} (t)$ but also modify the initial scheme (\ref{vis1}-\ref{bc1}), taking into account terms of the second order in the surface elevation $\xi(x)$.

Next, we will assume that $z_0 \gg l$ and therefore, as we explained earlier, the statistical properties of velocity fluctuations are insensitive to the fine details of the roughness statistics and the characteristic length scale in which correlation between fluctuations decays is of the order of the distance to the wall $z_0$. The expression (\ref{eq:delta_z}) is correct until $U_0 t \ll z_0 \sqrt{U_0/u}$, where $u$ is characteristic magnitude of velocity fluctuations, see Eq.~(\ref{covariance}). Otherwise, one has to take into account the contribution to $\delta z(t)$ produced by the second-order term $\rho_{2z}(t)$. By using expression (\ref{eq:delta_z}) and averaging over the roughness statistics, we obtain $\langle \delta z (t) \rangle = 0$ and
\begin{equation}\label{eq:dispersion}
\langle \delta z (t)^2 \rangle = \frac{1}{U_0^2} \int\limits_0^{U_0 t} d \zeta \int\limits_{-\zeta}^{\zeta} d \eta \, \langle u_z(0,z_0) u_z(\eta,z_0)\rangle.
\end{equation}
By substituting expression (\ref{Rzz2d}) and $U_0=s z_0$, for the 2D fluid flow over the rough plane we finally obtain
\begin{equation}\label{eq:answer}
  \langle \delta z (t)^2 \rangle = \frac{A}{\pi z_0} \frac{(st)^2}{(st)^2+4}, \quad st \ll \sqrt{U_0/u}.
\end{equation}
It means that the spatial dispersion in the wall-normal direction sufficiently quickly reaches the value $\langle \delta z^2 \rangle_{lim} = A/(\pi z_0)$ and no longer changes with time.
As $h\ll l\ll z_0$ and $A\sim h^2l$, we obtain the estimate $\langle \delta z^2 \rangle_{lim}\sim h^2l/z_0\ll z_0^2$.
 Therefore, the roughness-induced velocity fluctuations do not lead to the enhancement of diffusion of the passive scalar in the direction perpendicular to the wall at least in the leading order approximation considered here.

In principle, expression (\ref{eq:dispersion}) can be used to estimate the asymptotic value $\langle \delta z^2 \rangle_{lim}$ at any distance $z_0$ to the wall and for any geometry considered in this paper. The main contribution to the integrals comes from the scale of the correlation length of velocity fluctuations $l_{corr} = \mathrm{max} (z_0, l)$ and then
\begin{equation}\label{eq:estimation}
  \langle \delta z^2 \rangle_{lim} \sim \frac{\langle u_z^2 \rangle}{U_0^2 } l_{corr}^2.
\end{equation}
By using expression (\ref{covariance}), one can check that the estimation correctly reproduces the asymptotic value of dispersion for the 2D flow over rough wall in the far-field region. Note that for the 3D flow in the same geometry one can obtain $\langle \delta z^2 \rangle_{lim} \sim A/z_0^2$, $z_0 \gg l$, where $A \sim h^2 l^2$ and we have used expression (\ref{eq:covariance_3d}).

\section{Conclusion}

We investigated the influence of stochastic boundary roughness on the stationary Stokes flow of incompressible fluid in the limit when the typical height $h$ of the random asperities of the boundary is much smaller than their transversal size $l$. The focus of the paper is on the statistical properties of the spatially random flow component  produced in the presence of roughness. Being small compared with the mean flow far enough from the boundaries, the fluctuations in fluid velocity can be treated perturbatively there.

First, we considered the fluid flow in a half-space over flat imperfect wall in both $2$ and $3$ spatial dimensions.
While the near-wall properties of the fluid velocity fluctuations are sensitive to the statistics of wall roughness,  universal behavior emerges in the far field. Namely, as long as the variance of surface heights is finite, the velocity fluctuations at the distances $z \gg l$ from the wall possess Gaussian statistics. The far-field intensity of fluctuations as a function of distance from the wall exhibits a power-law decay with a dimension-dependent exponent which is insensitive to the details of the roughness statistics. In addition, we calculated the two-point correlation function and the energy spectrum of the velocity fluctuations revealing the power-law decay of spatial correlations.

Next, we considered the $2d$ and $3d$ flow in the plane channel between two parallel walls, one of which is rough, while another is perfectly flat.
Particular attention was paid to the energy spectrum of velocity fluctuations as a function of the distance to the rough wall.
 The analytical expression (\ref{eq:spectr2d_calc}) for the energy spectrum  is valid for any relation between the correlation length $l$ of imperfect wall roughness and the distance $d$ between the channel walls.
The main difference in comparison with the half-space geometry is the suppression of the large-scale velocity fluctuations inside the channel.

Finally, we considered the transport of passive scalar in the steady velocity field with a relatively weak random component generated by the wall imperfections. It turns out that the spatial dispersion of the passive scalar in the wall-normal direction reaches some value given by expression (\ref{eq:estimation}) on the time scale of the order of the inverse shear rate of the mean flow, and no longer changes with time. Therefore the random surface roughness does not lead to the enhancement of diffusion of the passive scalar in the direction perpendicular to the wall at least in the second order with respect to the roughness amplitude. Qualitatively, this result is
explained by the fact that Lagrangian trajectories of the laminar flow in the channel with small-amplitude wall roughness   do not diverge as the streamwise coordinate grows.


We hope that the obtained results will motivate further work on the impact of the random roughness on microfluidic flow
behavior. Various experimental tools  have been  successfully applied for roughness characterization, see \citet{Yu_1996,Yilbas_1999,Bhushan_2001,Duparre_2002,Zhu_2004,Kuo_2010,Ren_2011}. Laminar velocity field inside the microfluidic channels with micro/nano scale surface roughness can be measured with particle image velocimetry technique \citep{Silva_2008,Jaeger_2012,Ren_2013}. That encourage us to think that experimental test of our findings is the matter of near future.


We thank S. S. Vergeles for valuable discussions. This work was supported by the Russian Science Foundation, Grant No. 14-22-00259.

\appendix
\section{Characteristic Function}\label{appA}

By using the functional methods we can express the characteristic function through the integration over different realizations of the surface elevation $\xi(x)$
\begin{equation}
  G (\vec k) = \int {\cal D \xi Q(\xi)} e^{i \vec k \vec u[\xi]},
\end{equation}
where the velocity $\vec u [\xi]$ should be taken at some fixed point. Without the loss of generality we assume that this point has coordinates $(0,z)$. We also assume that $z \gg l$ and it means that the random heights $\xi(x)$ of surface asperities at different points of the wall can be considered as completely independent from each other.

To illustrate the functional integration we will use the discretization. We split the wall surface into many small pieces and then the streamwise coordinate takes discrete values $x_k = k \varepsilon$, where k is an integer number and $\varepsilon$ is the discretization step, $z \gg \varepsilon \gtrsim l$. For the characteristic function we obtain
\begin{equation}
 G (\vec k) = \prod_k \int d \xi_k P(\xi_k) e^{i \varepsilon \alpha (x_k) \xi_k},\quad \alpha (x_k) = \frac{2sz}{\pi} \frac{x_k(zk_z - x_k k_x)}{(x_k^2+z^2)^2},
\end{equation}
and here $\xi_k = \xi (x_k)$ are independent random variables, which have the same probability distribution $P$. By expanding the exponential term in a Taylor series, in the main order with respect to the parameter $l/z \ll 1$ we obtain:
\begin{eqnarray}
\nonumber
&\displaystyle G (\vec k) \simeq \prod_k \left( 1 - \frac{\langle \xi_k^2 \rangle \alpha^2(x_k) \varepsilon^2}{2} \right)= 1-\frac{\langle \xi^2\rangle\varepsilon^2}{2}\sum_k \alpha^2(x_k) + &\\
 \nonumber
&\displaystyle +\frac{\langle \xi^2\rangle^2\varepsilon^4}{4}\sum_{k_1\ne k_2}\alpha^2(x_{k_1})\alpha^2(x_{k_2})
 -\frac{\langle \xi^2\rangle^3\varepsilon^6}{8}\sum_{k_1\ne k_2\ne k_3}\alpha^2(x_{k_1})\alpha^2(x_{k_2})\alpha^2(x_{k_3})+ \dots = &\\
 \nonumber
&\displaystyle =1+ \sum_{n=1}^{\infty}\frac{(-1)^n}{2^n}\langle \xi^2\rangle^n\varepsilon^{2n}\sum_{k_1\ne \dots \ne k_n}\alpha^2(x_{k_1})\dots \alpha^2(x_{k_n})\simeq &\\
&\displaystyle \simeq \sum_{n=0}^{\infty}\frac{(-1)^n}{n!}\left(\frac{\varepsilon\langle \xi^2\rangle}{2}\int\limits_{-\infty}^{+\infty}dx \, \alpha^2(x)\right)^n= \exp \left[ - \frac{\varepsilon \langle \xi^2 \rangle}{2} \int\limits_{-\infty}^{\infty} dx \, \alpha^2(x) \right]. &
\end{eqnarray}
The quantity $\varepsilon \langle \xi^2 \rangle$ plays a role of the area under the curve $F(r)$ and should be substituted by $A$, according to the definition (\ref{eq:A}). After the integration one finds the answer (\ref{char_function}).

\end{document}